\documentclass[12pt]{article}
\usepackage{amssymb}
\usepackage{epsfig}

\catcode`\@=11
\def\numberbysection{\@addtoreset{equation}{section}
        \def\theequation{\thesection.\arabic{equation}}}
\def\half{\frac{1}{2}}
\def\beq{\begin{equation}}
\def\eeq{\end{equation}}
\numberbysection
\begin{document}
\begin{titlepage}
\begin{center}
\hfill DFF  03/03/03 \\
\vskip 1.in {\Large \bf Matrix model for noncommutative gravity
and gravitational instantons} \vskip 0.5in P. Valtancoli
\\[.2in]
{\em Dipartimento di Fisica, Polo Scientifico Universit\'a di Firenze \\
and INFN, Sezione di Firenze (Italy)\\
Via G. Sansone 1, 50019 Sesto Fiorentino, Italy}
\end{center}
\vskip .5in
\begin{abstract}
We introduce a matrix model for noncommutative gravity, based on
the gauge group $U(2) \otimes U(2)$. The vierbein is encoded in a
matrix $Y_{\mu}$, having values in the coset space $U(4)/ (U(2)
\otimes U(2))$, while the spin connection is encoded in a matrix
$X_\mu$, having values in $U(2) \otimes U(2)$. We show how to
recover the Einstein equations from the $\theta \rightarrow 0$
limit of the matrix model equations of motion. We stress the
necessity of a metric tensor, which is a covariant representation
of the gauge group in order to set up a consistent second order
formalism. We finally define noncommutative gravitational
instantons as generated by $U(2) \otimes U(2)$ valued
quasi-unitary operators acting on the background of the Matrix
model. Some of these solutions have naturally self-dual or
anti-self-dual spin connections.
\end{abstract}
\medskip
\end{titlepage}
\pagenumbering{arabic}
\section{Introduction}

Unification of noncommutative geometry with gravity theories is a
very challenging goal for a theoretical physicist
\cite{1}-\cite{12}. Till now only gauge theories are proved to be
consistent after deforming the ordinary product of fields into a
noncommutative star product.

Gravity theories are usually constructed by requiring either
diffeomorphism invariance or local Lorentz invariance. Since the
Moyal star product breaks global Lorentz invariance in gauge
theories, it also breaks diffeomorphism invariance in a gravity
theory. Therefore in the noncommutative case it is possible to
preserve only local Lorentz invariance, with the gauge group
extended to $U(1,d-1)$ instead of $SO(1,d-1)$, in order that the
gauge transformations multiplied with the star product are closed
between them.

This approach has still many problems, mainly because the metric
becomes complex, and the antisymmetric part of the metric may have
nonphysical propagating modes \cite{3}-\cite{5}. It is however
worth to explore all the consequences that such a new theory can
give, before taking an opinion about it. In this paper we attempt
to give a more solid construction of noncommutative gravity theory
by introducing a new matrix model based on the $U(2) \otimes U(2)$
group ( for Euclidean gravity ) or $U(1,1) \otimes U(1,1)$ for the
$U(2,2)$ case.

In this respect the first-order formalism, based on the vierbein
and spin connection, turns out to be superior to the second order
formalism, since it permits the definition of a matrix model,
without the necessity of inverting the metric, which would be a
rather difficult obstacle.

The vierbein is encoded in a matrix $Y_\mu$, having values in the
coset space $ \frac{U(4)}{U(2) \otimes U(2)} $, while the spin
connection is encoded in a matrix $X_\mu$, having values in the
gauge group $U(2) \otimes U(2)$. The action is a $4$-form, thus
preventing the use of any metric, and the corresponding equations
of motion are proved to be a natural generalization of the
Einstein equations, provided a certain separation between odd
powers in $\theta$ and even powers in $\theta$ is made ( as done
in Ref. \cite{9} and Ref. \cite{11} ).

We confirm the necessity of a complex metric tensor, but we point
out a property which was not discussed before, i.e. the metric
tensor cannot be defined to be invariant under local Lorentz
invariance, but at most it can be a covariant representation of
the gauge group. We believe that the introduction of a multiplet
of metric tensors is necessary to set up a consistent second order
formalism. Each component of this multiplet has no direct physical
meaning, since it can be mixed with the other components by a
gauge redefinition of the vierbein and spin connection.

In the last part of the article, we attack the problem of defining
noncommutative gravitational instantons \cite{13}-\cite{17}. In
this sense, our formalism based on the Matrix model approach turns
out to be fruitful, since then the finite action solutions of the
equations of motion are generated by quasi-unitary operators, as
it has been successfully found in the Yang-Mills case
\cite{20}-\cite{25}. The ultimate source of our quasi-unitary
operator is a projector, with values in $U(2)_{\rm left} \otimes
U(2)_{\rm right}$. We find that if the projector is restricted to
be pure left or pure right, then the corresponding spin connection
is self-dual or anti-self-dual, an important property which can
make the bridge between our definition of gravitational instantons
and the classical standard definition. However our class of
solutions is more general than only self-dual ones. We finally
outline how to construct examples of quasi-unitary operators
giving rise to their sources, the $U(2) \otimes U(2)$ projectors.
An important tool to construct these examples is the use of
duality in noncommutative theory \cite{18}-\cite{19}.

\section{Matrix model for noncommutative gravity}

We are going to introduce a gravity theory on a noncommutative
plane defined by the commutators:

\beq [ \hat{x}_i, \hat{x}_j ] = i \theta_{ij} \ \ \ \ \ \ \ \ \
det |\theta_{ij}| \neq 0 . \label{21}\eeq

We remember that instead of using the commutation relations in
full generality, it is possible to reduce them in a diagonal form
as follows

\begin{eqnarray}
& \ & [ \hat{x}_1, \hat{x}_2 ] = i \theta_1 \nonumber \\
& \ & [ \hat{x}_3, \hat{x}_4 ] = i \theta_2 ,\label{22}
\end{eqnarray}
which is basically solved by two types of raising and lowering
oscillator operators.

With the experience in the Yang-Mills case \cite{25}, we attack
the problem of noncommutative gravity. Our proposal is based on
gauging the local Lorentz symmetry, extended to $U(2) \otimes
U(2)$ for consistency.

Firstly we define two types of matrices $X_\mu$, $Y_\mu$, where
$X_\mu$ is, at least in the Euclidean case, a hermitian matrix
with values in the group $U(2) \otimes U(2)$, while $Y_\mu$ is a
hermitian matrix with values in the coset space $\frac{ U(4) }{
U(2) \otimes U(2) }$. Then we write the Einstein action for the
noncommutative case as

\begin{eqnarray} S_E & = & \beta_E \ Tr [ \gamma_5 \
\epsilon^{\mu\nu\rho\sigma} Y_\mu Y_\nu ( [ X_\rho , X_\sigma ] -
i \theta_{\rho\sigma}^{-1} ) ] \nonumber \\
  Y^{\dagger}_\mu & = & Y_\mu \ \ \ \ \ \ \ \ X^{\dagger}_\mu = X_\mu
\ \ \ \ \ \ \ \gamma_5 = \gamma_0 \gamma_1 \gamma_2 \gamma_3 .
\label{23}
\end{eqnarray}

This action can be derived as a part of a more symmetric action (
inspired by Ref. \cite{2} and Ref. \cite{9} ),  based on two
$U(4)$ fields $X_\mu^{\pm}$

\beq X_\mu^{\pm} = X_\mu \pm Y_\mu , \label{24} \eeq

given by the sum of two similar contributions:

\begin{eqnarray}
S & = & \beta \ Tr [ \gamma_5 \ \epsilon^{\mu\nu\rho\sigma} ( [
X_\mu^{+}, X_\nu^{+} ] - i \theta^{-1}_{\mu\nu} ) ( [ X_\rho^{+},
X_\sigma^{+} ] - i \theta^{-1}_{\rho\sigma} ] \nonumber \\
& + & \beta \ Tr [ \gamma_5 \ \epsilon^{\mu\nu\rho\sigma} ( [
X_\mu^{-}, X_\nu^{-} ] - i \theta^{-1}_{\mu\nu} ) ( [ X_\rho^{-},
X_\sigma^{-} ] - i \theta^{-1}_{\rho\sigma} ] .\label{25}
\end{eqnarray}

With the trick of adding the second term, the odd powers in
$Y_\mu$ all vanish and we are left with three terms; the term with
zero powers of $Y_\mu$ is a pure topological term, representing
noncommutative topological gravity, the term with two powers of
$Y_\mu$ reproduces the action (\ref{23}) which is the starting
point of our article, and a term proportional to the square of the
torsion $T_{\mu\nu}$, defined as

\beq T_{\mu\nu} = [ X_\mu, Y_\nu ] - [ X_\nu, Y_\mu ], \label{26}
\eeq

while the term with four powers of $Y_\mu$ gives rise to the
cosmological constant term.

Being the dependence from the torsion $T_{\mu\nu}$ quadratic, it
is possible to set it equal to zero, because the variation of the
other terms are consistent with this choice.

Neglecting the topological term and the cosmological constant term
we can continue to discuss the action (\ref{23}) as the basis for
noncommutative gravity.

In this paper we mainly discuss the $U(4)$ and $U(2,2)$ cases for
simplicity, since we are mainly interested to introduce the
noncommutative version of gravitational instantons, and we need to
work with the Euclidean case. In the Euclidean case $U(4)$, the
gamma matrices satisfy the hermitian condition

\beq \gamma^{\dagger}_a = \gamma_a \ \ \ \ \ \ \ \
\gamma_5^{\dagger} = \gamma_5 . \label{27}\eeq

The matrix $Y_\mu$ can be developed in terms of basic $U(1)$
valued matrices :

\beq Y_\mu = e^a_\mu \gamma_a + i f^a_\mu \gamma_a \gamma_5 .
\label{28} \eeq

The hermitian condition of the $Y_\mu$ is reflected on a hermitian
condition on the component vierbeins as follows:

\begin{eqnarray}
& \ & ( e^a_\mu )^{\dagger} = e^a_\mu \nonumber \\
& \ & ( f^a_\mu )^{\dagger} = f^a_\mu . \label{29}
\end{eqnarray}

It is possible to define left and right combinations of the
vierbeins as follows:

\beq e^{\pm a}_{\mu} = e^a_\mu \pm i f^a_\mu \ \ \ \ \ ( e^{+
a}_\mu )^{\dagger} = e^{- a}_\mu , \label{210} \eeq

which are related by hermitian conjugation. These matrices have
the nice property to be closed under the $U(2) \otimes U(2)$ gauge
transformations. In fact a left-right decomposition can be made at
the level of the $Y_\mu$ defining

\begin{eqnarray}
Y_\mu & = & Y_\mu^{+} + Y_{\mu}^{-} \nonumber \\
Y_{\mu}^{\pm} & = & e^{\pm a }_{\mu} \gamma_a \left( \frac{1 \pm
\gamma_5}{2} \right) . \label{211}
\end{eqnarray}

In terms of $Y^{\pm}_\mu$ the action can be rewritten as

\beq S_E = - \beta_E \ Tr [ \epsilon^{\mu\nu\rho\sigma} (
Y_{\mu}^{+} Y_{\nu}^{-} + Y_{\mu}^{-} Y_{\nu}^{+}) ( [ X_\rho,
X_\sigma ] - i \theta^{-1}_{\rho\sigma} ) ] . \label{212} \eeq

Note that since $Y^{+}_\mu Y^{+}_\nu = 0 $, it is not possible to
define an action containing only left combinations of the
vierbein, but both left and right vierbeins are required to make
$S_E$ hermitian.

The matrix $X_\mu$ can be developed in terms of the basic
matrices:

\beq X_\mu = \hat{p}_\mu + \omega_\mu^1 + \omega_\mu^5 \gamma^5 +
i \omega_{\mu}^{ab} \gamma_{ab} \ \ \ \ \ \ \ \gamma_{ab} = \half
[ \gamma_a, \gamma_b ] , \label{213} \eeq

where the component matrices are all hermitian by construction:
\beq  \hat{p}_\mu^{\dagger} = \hat{p}_\mu \ \ \ \ \omega_\mu^{1 \
\dagger} = \omega_\mu^1 \ \ \ \ \ \omega_\mu^{5 \ \dagger} =
\omega_\mu^5 \ \ \ \ \ \ \ \omega_{\mu}^{ab \ \dagger} =
\omega_{\mu}^{ab} . \label{214} \eeq

We have introduced the distinction between $\hat{p}_\mu$ and
$\omega_{\mu}^1$ since in a matrix model for noncommutative
gravity it is necessary to separate the background from the
fluctuations \cite{25}. In general, the background for the matrix
model of noncommutative gravity will be defined as

\begin{eqnarray}
X_\mu & = & \hat{p}_\mu        \ \ \ \ \ \  \hat{p}_\mu = -
\theta_{\mu\nu}^{-1} \hat{x}_\nu \nonumber \\
Y_\mu & = & \delta_\mu^a \gamma_a , \label{215}
\end{eqnarray}

and the fluctuations are generated by the matrices

\beq \omega_\mu^1, \ \omega_\mu^5, \ \omega^{ab}_\mu, \ e^a_\mu -
\delta^a_\mu, \ f^a_\mu . \label{216} \eeq

The background is chosen to satisfy the equations of motion of the
matrix model and to introduce the noncommutative coordinates as
noncommutative analogues of the concepts of derivatives of the
fields, as we normally do in the Yang-Mills case. Therefore the
commutation relation , for example, of the background
$\hat{p}_\mu$ with the matrix $\omega_\nu^1$ is equivalent, at a
level of the corresponding symbol, to a derivative action:

\beq [ \hat{p}_\mu, \omega_\nu^1 ] \rightarrow \partial_\mu
\omega_\nu^1 . \label{217} \eeq

Moreover the operator products are transformed into star products
of the corresponding symbols ( through the Weyl map, see Ref.
\cite{18} for details ).

In general, the matrix model is built on the gauge invariance

\begin{eqnarray}
X_\mu & \rightarrow & U^{-1} X_\mu U  \nonumber \\
Y_\mu & \rightarrow & U^{-1} Y_\mu U , \label{218}
\end{eqnarray}

where $U$ is the gauge transformation of $U(2) \otimes U(2)$;
these types of transformations reproduce, in the commutative
limit, the standard local Lorentz transformations for the vierbein
and the spin connection. The gauge transformations of $U(2)
\otimes U(2)$ are defined from the generators $ 1, \gamma_5 ,
\gamma_{ab}$ and obey the unitary condition:

\beq U^{\dagger} U = U U^{\dagger} = 1 . \label{219} \eeq

Therefore introducing the anti-hermitian matrix $\Lambda$

\beq U = \exp [ \Lambda ]  \ \ \ \ \Lambda^{\dagger} = - \Lambda
\label{220} \eeq

$\Lambda$ can be developed in terms of basic gauge parameters:

\beq \Lambda = i \Lambda_0 + i \Lambda_5 \gamma_5 + \Lambda^{ab}
\gamma_{ab} , \label{221} \eeq

where the component parameters are all hermitian matrices:

\beq \Lambda_0^{\dagger} = \Lambda_0 \ \ \ \ \ \Lambda_5^{\dagger}
= \Lambda_5 \ \ \ \ \ \ \Lambda^{ab \ \dagger} = \Lambda^{ab} .
\label{222} \eeq

What does it change in the $U(2,2)$ scenario ? The action is
always the same (\ref{23}) with $\gamma_5 = \gamma_0 \gamma_1
\gamma_2 \gamma_3$ ( see Appendix ). The matrix $Y_{\mu}$ is a
matrix with values in the coset space $\frac{U(2,2)}{U(1,1)
\otimes U(1,1)}$, and the matrix $X_\mu$ has values in $U(1,1)
\otimes U(1,1)$.

To define the hermitian conjugation of these matrices, let us
recall that $\Gamma_0 = \gamma_0 \gamma_1$ is the hermitian
conjugation matrix for the gamma matrices in the $U(2,2)$ case:

\begin{eqnarray}
\gamma^2_0 & = & \gamma^2_1 = - \gamma^2_2 = - \gamma^2_3 = 1 \ \
\ \ \ \ \ \{ \gamma_a , \gamma_b \} = 2 \eta_{ab} \ \ \ \ ( + + -
- ) \nonumber \\
\gamma^{\dagger}_a & = & \Gamma_0 \gamma_a \Gamma_0 \ \ \ \ \ \ \
\gamma_5^{\dagger} = \gamma_5 \ \ \ \ \  \gamma_5 \Gamma_0 =
\Gamma_0 \gamma_5 . \label{223} \end{eqnarray}

Then we define the hermitian conjugation for the matrices $Y_\mu,
X_\mu$ as

\begin{eqnarray}
Y^{\dagger}_\mu & = & \Gamma_0 Y_\mu \Gamma_0 \nonumber \\
X^{\dagger}_\mu & = & - \Gamma_0 X_\mu \Gamma_0 . \label{224}
\end{eqnarray}

Again $Y_\mu$ can be developed in terms of the basic components as
follows:

\begin{eqnarray}
Y_\mu & = & e^a_\mu \gamma_a + i f^a_\mu \gamma_a \gamma_5
\nonumber \\
(e^a_\mu)^{\dagger} & = & e^a_\mu \ \ \ \ \ (f^a_\mu)^{\dagger} =
f^a_\mu , \label{225}
\end{eqnarray}

and $X_\mu$

\beq X_\mu = \hat{p}_\mu + \omega_\mu^1 + \omega_\mu^5 \gamma_5 +
i \omega_{\mu}^{ab} \gamma_{ab} , \label{226} \eeq

where all the components are hermitian.

In general the $U(2,2)$ matrix model is built on the gauge
invariance

\begin{eqnarray}
X_{\mu} & \rightarrow & U^{-1} X_{\mu} U \nonumber \\
Y_{\mu} & \rightarrow & U^{-1} Y_{\mu} U , \label{227}
\end{eqnarray}

where $U$ is a gauge transformation of $U(1,1) \otimes U(1,1)$.
The gauge transformations of this group are again defined from the
generators $ 1, \gamma_5, \gamma_{ab}$ and obey the condition

\begin{eqnarray}
U^{\dagger} \Gamma_0 U = \Gamma_0 \nonumber \\
U \Gamma_0 U^{\dagger} = \Gamma_0 . \label{228}
\end{eqnarray}

This reality condition assures that the matrix $X_\mu$, once that
it is gauge transformed, respects again the hermitian condition
(\ref{224}).

By defining $U = \exp [ \Lambda ]$ it follows that

\beq \Lambda^{\dagger} = \Gamma_0 \Lambda \Gamma_0 . \label{229}
\eeq

The matrix $\Lambda$ can be developed in terms of the basic
parameters:

\beq \Lambda = i \Lambda_0 + i \Lambda_5 \gamma_5 + \Lambda^{ab}
\gamma_{ab} \label{230} \eeq

with all hermitian components.

Finally let us discuss the decomposition in $U(2)_{\rm \ left}$
and $U(2)_{\rm \ right}$ of the Euclidean $U(2) \otimes U(2)$
gauge transformations. This decomposition is obtained by requiring
that the generators of $U(2)_{\rm \ left}$ are given by

\beq ( 1, \gamma_{ab} ) \left( \frac{1 + \gamma_5}{2} \right).
\label{231} \eeq

The application of the projector operator $\frac{1+\gamma_5}{2}$
on $\gamma_{ab}$ reduces the number of generators from six to
three. To see this property in detail we recall the identity:

\beq \gamma_5 \gamma_{ab} = - \half \epsilon_{abcd} \gamma_{cd}.
\label{232} \eeq

Therefore the generators of $SU(2)_{\rm \ left}$ read

\beq M_{ab} = \half [ \gamma_{ab} - \half \epsilon_{abcd}
\gamma_{cd} ] .\label{233} \eeq

Since $ M_{ab} = - \half \epsilon_{abcd} M_{cd} $ is
anti-self-dual, the only independent generators are three

\beq M_{0i} = \half [ \gamma_{0i} - \half \epsilon_{ijk}
\gamma_{jk} ] = - i \left( \begin{array}{cc} \sigma_i & 0 \\
0 & 0 \end{array} \right) . \label{234} \eeq

Analogously the generators of $U(2)_{\rm \ right}$ are given by

\beq ( 1, \gamma_{ab} ) \left( \frac{1 - \gamma_5}{2} \right) ,
\label{235} \eeq

where

\beq N_{ab} = \frac{1-\gamma_5}{2} \gamma_{ab} = \half [
\gamma_{ab} + \half \epsilon_{abcd} \gamma_{cd} ] \label{236} \eeq

is self-dual

\beq N_{ab} = \half \epsilon_{abcd} N_{cd} . \label{237} \eeq

Therefore the only independent generators are

\beq N_{0i} = \half [ \gamma_{oi} + \half \epsilon_{ijk}
\gamma_{jk} ] = i \left( \begin{array}{cc} 0 & 0 \\ 0 & \sigma_i
\end{array} \right) . \label{238} \eeq

What happens in the $U(1,1) \otimes U(1,1)$ case ?

The generators of $U(1,1)_{\rm \ left}$ are given by

\beq ( 1, \gamma_{ab} ) \frac{1+ \gamma_5}{2} , \label{239} \eeq

where now the identity  (\ref{232}) reads $ \gamma_5 \gamma_{ab} =
- \half \epsilon_{abcd} \gamma^{cd} $.

One can always define the generators of $U(1,1)_{\rm \ left}$ and
$U(1,1)_{\rm \ right}$ as

\begin{eqnarray}
M_{ab} & = & \half [ \gamma_{ab} - \half \epsilon_{abcd}
\gamma^{cd} ] \nonumber \\
N_{ab} & = & \half [ \gamma_{ab} + \half \epsilon_{abcd}
\gamma^{cd} ] , \label{240}
\end{eqnarray}

where now for $ su(1,1)_{\rm \ left}$

\begin{eqnarray}
M_{01} & = & \half [ \gamma_{01} - \gamma_{23} ] = -i \left(
\begin{array}{cc} \sigma_1 & 0 \\ 0 & 0 \end{array} \right) \nonumber
\\
M_{02} & = & \half [ \gamma_{02} + \gamma_{31} ] =  \left(
\begin{array}{cc} \sigma_2 & 0 \\ 0 & 0 \end{array} \right) \nonumber
\\
M_{03} & = & \half [ \gamma_{03} + \gamma_{12} ] =  \left(
\begin{array}{cc} \sigma_3 & 0 \\ 0 & 0 \end{array} \right)
\label{241}
\end{eqnarray}

and for $su(1,1)_{\rm \  right}$

\begin{eqnarray}
N_{01} & = & \half [ \gamma_{01} + \gamma_{23} ] = i \left(
\begin{array}{cc} 0 & 0 \\ 0 & \sigma_1 \end{array} \right) \nonumber
\\
N_{02} & = & \half [ \gamma_{02} - \gamma_{31} ] =  - \left(
\begin{array}{cc} 0 & 0 \\ 0 & \sigma_2 \end{array} \right) \nonumber
\\
N_{03} & = & \half [ \gamma_{03} - \gamma_{12} ] =  - \left(
\begin{array}{cc} 0 & 0 \\ 0 & \sigma_3 \end{array} \right).
\label{242}
\end{eqnarray}

Let us define the chiral decomposition of the gauge transformation
as

\beq U = \frac{1+\gamma_5}{2} U_L + \frac{1-\gamma_5}{2} U_R .
\label{243} \eeq

The left part $U_L$ must obey the hermitian condition

\begin{eqnarray} & \ &  U^{\dagger}_L \frac{1+\gamma_5}{2} \Gamma_0 U_L =
\frac{1+\gamma_5}{2} \Gamma_0 \nonumber \\
& \ & \frac{1+\gamma_5}{2} \Gamma_0 = i \sigma_1 \left(
\begin{array}{cc} 1 & 0 \\ 0 & 0 \end{array} \right) . \label{244}
\end{eqnarray}

Therefore reducing eq. (\ref{244}) to the upper $2 \times 2$
subspace, we find

\beq U^{\dagger}_L \sigma_1 U_L = \sigma_1 \label{245} \eeq

and analogously for $U_R$

\beq U^{\dagger}_R \sigma_1 U_R = \sigma_1 , \label{246} \eeq

which are completely equivalent to the usual $U(1,1)$ condition

\beq U^{\dagger} \sigma_3 U = \sigma_3 . \label{247} \eeq

By defining

\beq U_{L,R} = \exp [ \Lambda_{L,R} ] \ \ \ \ \ \
\Lambda^{\dagger}_{L,R} = -\sigma_1 \Lambda_{L,R} \sigma_1 ,
\label{248} \eeq

it follows that $\Lambda $ can be expanded  as

\beq \Lambda = i \Lambda_0 + i \Lambda_1 \sigma_1 + \Lambda_2
\sigma_2 + \Lambda_3 \sigma_3 , \label{249} \eeq

where all the components are hermitian.

\section{Equations of motions}

Let us discuss the equations of motion of the matrix model. Since
there are two independent matrices, we need to vary the action
with respect to $\delta X_\mu$ and $\delta Y_\mu$ independently,
therefore obtaining two types of equations of motion.

The first one is due to the variation with respect to $\delta
X_\mu$

\beq \delta S = Tr [ \gamma_5 Y_\mu Y_\nu ( \delta X_\rho X_\sigma
+ X_\rho \delta X_\sigma ) \epsilon^{\mu\nu\rho\sigma} ]
\label{31} \eeq

that is vanishing if the following tensor is null

\beq T_{\mu\nu} = [ X_\mu, Y_\nu ] - [ X_\nu, Y_\mu ] = 0
\label{32}\eeq

i.e. it is equivalent to the condition of null torsion.

The other equation of motion is obtained by varying with respect
to $\delta Y_\mu$

\beq \delta S = Tr [ \gamma_5 \epsilon^{\mu\nu\rho\sigma} ( \delta
Y_\mu Y_\nu + Y_\mu \delta Y_\nu ) R_{\rho\sigma } ] \label{33}
\eeq

where

\beq R_{\mu\nu} = [ X_\mu, X_\nu ] - i \theta_{\mu\nu}^{-1}
\label{34} \eeq

and it is vanishing if the following condition is met

\beq \epsilon^{\mu\nu\rho\sigma} \{ Y_\nu, R_{\rho\sigma} \} = 0 .
\label{35} \eeq

The two equations of motion (\ref{32}) and (\ref{35}) are not
completely independent, since it exists, at the noncommutative
level, an identity similar to the covariant conservation of the
tensor $ G_{\mu\nu}$

\beq D_\mu G^{\mu\nu} = 0 \ \ \ \ \   G^{\mu\nu} = R^{\mu\nu}_E -
\half g^{\mu\nu} R_E \label{36} \eeq

where we have distinguished $R_E^{\mu\nu}$, the Einstein Ricci
tensor, from the antisymmetric matric $R_{\mu\nu}$ introduced
before.

In fact, let us evaluate

\beq \epsilon^{\mu\nu\rho\sigma} [ X_\mu, \{ Y_\nu, R_{\rho\sigma}
\} ] = 0 .\label{37} \eeq

We will prove that it corresponds to a trivial identity. It is
enough to observe that (\ref{37}) can be decomposed into a sum of
terms which are zero

\beq \epsilon^{\mu\nu\rho\sigma} [ X_\mu, Y_\nu ] = 0 \label{38}
\eeq

because of the null torsion condition, and

\beq \epsilon^{\mu\nu\rho\sigma} [ X_\mu, R_{\rho\sigma} ] = 0
\label{39} \eeq

because of the Jacobi identity.

Let us write the equations of motion (\ref{32}) and (\ref{35}) in
components, to recognize the usual form of the Einstein equations
in the commutative limit.

By introducing the parameterizations (\ref{28}) and (\ref{213}) of
the Euclidean case we find that $T_{\mu\nu} = 0 $ is equivalent to

\begin{eqnarray}
& \ & [ \hat{p}_\mu, e^a_\nu ] + [ \omega^1_\mu, e^a_\nu ] - i \{
\omega^5_\mu, f^a_\nu \} + 2i \{ \omega^{ab}_{\mu}, e^b_\nu \} -
\epsilon_{abcd} [ \omega^{bc}_\mu, f^d_\nu ] \nonumber \\
& \ & = ( \mu \leftrightarrow \nu ) \nonumber \\
& \ & [ \hat{p}_\mu, f^a_\nu ] + [ \omega^1_\mu, f^a_\nu ] + i \{
\omega^5_\mu, e^a_\nu \} + 2i \{ \omega^{ab}_{\mu}, f^b_\nu \} +
\epsilon_{abcd} [ \omega^{bc}_\mu, e^d_\nu ] \nonumber \\
& \ & = ( \mu \leftrightarrow \nu ) . \label{310}
\end{eqnarray}

Therefore it is not possible to make the two vierbeins $e^a_\mu$
and $f^a_\mu$ proportional, because of the term proportional to
$\omega^5_\mu$ and $\epsilon_{abcd}$.

These two equations relate all the components of the spin
connection $ \omega^1_\mu, \omega^5_\mu, \omega^{ab}_\mu$ in terms
of the generic vierbeins ($ e^a_\mu, f^a_\mu $), treating them
independently.

To help intuition, it is possible to restrict the general
equations of motion (\ref{310}) such that the symbol corresponding
to the first operator equation contains only even powers of
$\theta$, while the symbol corresponding to the second equation
contains only odd powers of $\theta$ ( we refer to Ref. \cite{9}
for a detailed discussion on this point ).

This reduction requires that the symbols of the operators
$e^a_\mu, \omega^{ab}_\mu$ have an expansion in $\theta$ with only
even powers, while the symbols of the operators $f^a_\mu,
\omega^1_\mu, \omega^5_\mu$ have only odd powers. In this
scenario, the usual free torsion condition for $e^a_\mu$ is
recovered in the commutative limit since

\beq  [ \omega^1_\mu, e^a_\nu ] \sim  \{ \omega^5_\mu, f^a_\nu \}
\sim  [ \omega^{bc}_\mu, f^a_\nu ] \sim O ( \theta^2 ) ,
\label{311} \eeq

taking into account that the commutator of operators,
corresponding to the antisymmetric part of the star product, gives
another odd contribution to the powers of $\theta$, while the
anticommutator is even.

Let us analyze in the same scenario the other equation of motion
(\ref{35}) which should give rise to the usual Einstein equations.
Firstly let us compute the components of $R_{\mu\nu}$ as follows

\begin{eqnarray}
R_{\mu\nu} & = &[ X_\mu, X_\nu ] - i \theta^{-1}_{\mu\nu} =
\nonumber \\
& = & R^1_{\mu\nu} + R^5_{\mu\nu} \gamma^5 + R^{ab}_{\mu\nu}
\gamma_{ab} \nonumber \\
R^1_{\mu\nu} & = & [ \hat{p}_\mu , \omega^1_\nu ] - [ \hat{p}_\nu,
\omega^1_\mu ] + [ \omega^1_\mu, \omega^1_\nu ] +
[ \omega^5_\mu, \omega^5_\nu ] \nonumber \\
& = & -2 [ \omega^{ab}_\mu, \omega^{ba}_{\nu} ] \nonumber \\
R^5_{\mu\nu} & = & [ \hat{p}_\mu , \omega^5_\nu ] - [ \hat{p}_\nu,
\omega^5_\mu ] + [ \omega^1_\mu, \omega^5_\nu ] -
[ \omega^1_\nu, \omega^5_\mu ] \nonumber \\
& = & - \epsilon_{abcd} [ \omega^{ab}_\mu, \omega^{cd}_{\nu} ] \nonumber \\
R^{ab}_{\mu\nu} & = & i [ \hat{p}_\mu, \omega^{ab}_\nu ] - i [
\hat{p}_\nu, \omega^{ab}_{\mu} ] + i [ \omega^1_\mu,
\omega^{ab}_{\nu} ] - i [ \omega^1_\nu, \omega^{ab}_\mu ]
\nonumber \\
& - & \frac{i}{2} \epsilon_{abcd} ( [ \omega^5_\mu,
\omega^{cd}_\nu ] - [ \omega^5_\nu, \omega^{cd}_\mu ] ) + 2 \{
\omega^{bc}_\mu \omega^{ca}_\nu \} - 2 \{ \omega^{ac}_\mu
\omega^{cb}_\nu \} . \label{312}
\end{eqnarray}

Following the same reasoning done for the vierbein ( from now on
the distinction between odd and even powers of $\theta$ is always
intended true for the symbols of the corresponding operators ), it
is possible to restrict $R_{\mu\nu}$ such that $R^1_{\mu\nu},
R^5_{\mu\nu}$ contain only odd powers of $\theta$, while
$R_{\mu\nu}^{ab}$ contains only even powers of $\theta$, and
therefore in the commutative limit only $R^{ab}_{\mu\nu}$
survives.

Then we are ready to compute the equations of motion (\ref{35}):

\begin{eqnarray}
\epsilon^{\mu\nu\rho\sigma} [ \{ e^a_\nu, R^1_{\rho\sigma} \} + i
[ f^a_\nu, R^5_{\rho\sigma} ] + 2 [ e^b_\nu, R^{ba}_{\rho\sigma} ]
+ i \epsilon^{abcd} \{ f^d_\nu, R^{bc}_{\rho\sigma} \} ] = 0
\nonumber \\
\epsilon^{\mu\nu\rho\sigma} [ \{ f^a_\nu, R^1_{\rho\sigma} \} - i
[ e^a_\nu, R^5_{\rho\sigma} ] + 2 [ f^b_\nu, R^{ba}_{\rho\sigma} ]
- i \epsilon^{abcd} \{ e^d_\nu, R^{bc}_{\rho\sigma} \} ] = 0 .
\label{313}
\end{eqnarray}

Again it is not possible to make the two vierbeins proportional
because of the odd terms $R^5_{\rho\sigma}$ and $\epsilon^{abcd}$.

While the first equation of (\ref{313}) can be restricted to odd
powers of $\theta$, the second one can contain only even powers of
$\theta$. Since

\beq \{ f^a_\nu, R^1_{\rho\sigma} \} \sim [ e^a_\nu,
R^5_{\rho\sigma} ] \sim [ f^b_\nu, R^{ba}_{\rho\sigma} ] \sim
O(\theta^2) \label{314} \eeq

the only surviving term in the commutative limit is

\beq \epsilon^{\mu\nu\rho\sigma} \epsilon_{abcd} \{ e^d_\nu,
R^{bc}_{\rho\sigma} \} = 0 \label{315} \eeq

which is in fact completely equivalent to the usual Einstein
equations, where

\beq R^{bc}_{\rho\sigma} \sim \partial_\rho \omega^{bc}_\sigma -
\partial_\sigma \omega^{bc}_\rho + 2 \{ \omega^{cd}_{\rho} ,
\omega^{db}_{\sigma} \} - 2 \{ \omega^{bd}_{\rho} ,
\omega^{dc}_{\sigma} \} \label{316} \eeq

is the usual antisymmetric classical tensor.

\section{Gauge transformations of $Y_\mu$ and $X_\mu$ }

Before analyzing the gauge transformations of $Y_\mu$ and $X_\mu$,
we recall that the distinction between odd and even powers of
$\theta$ must be made also at a level of the gauge group $U(2)
\otimes U(2)$, to be consistent.

Recalling the results of Ref. \cite{11}, it is possible to reduce
the gauge symmetry from $U(2) \otimes U(2)$ to $SO(4)_{*}$, where
the gauge parameter $\Lambda$

\beq U = \exp [ \Lambda ] \ \ \ \ \ \ \ \Lambda = i \Lambda_0 + i
\Lambda_5 \gamma^5 + \Lambda^{ab} \gamma_{ab} \label{41} \eeq

has the following property, i.e. $\Lambda_0, \Lambda_5$ contains
only odd powers of $\theta$, while $\Lambda^{ab}$ contains only
even powers of $\theta$. This is a group property in the sense
that the commutator of two gauge parameters has the same
distinction.

We are now ready to analyze the gauge transformations of $Y_\mu$
and $X_\mu$. The vierbein $Y_\mu$ transforms under $U(2) \otimes
U(2)$ following the law

\beq Y_\mu \rightarrow U^{\dagger} Y_\mu U . \label{42} \eeq

At an infinitesimal level, it transforms as

\beq \delta Y_\mu = [ Y_\mu, \Lambda ] . \label{43} \eeq

By introducing the decomposition of $Y_\mu$ and $\Lambda$ in terms
of basic components we find

\begin{eqnarray}
\delta e^a_\mu & = & i [ e^a_\mu, \Lambda_0 ] + 2 \{ e^b_\mu,
\Lambda^{ba} \} - \{ f^a_\mu, \Lambda_5 \} + i \epsilon_{abcd} [
f^d_\mu, \Lambda^{bc} ] \nonumber \\
\delta f^a_\mu & = & i [ f^a_\mu, \Lambda_0 ] + 2 \{ f^b_\mu,
\Lambda^{ba} \} + \{ e^a_\mu, \Lambda_5 \} - i \epsilon_{abcd} [
e^d_\mu, \Lambda^{bc} ] . \label{44}
\end{eqnarray}

A rapid check shows that the first equation can be reduced to
contain, in the case of $SO(4)_{*}$ transformations, only even
powers of $\theta$, while the second equation only odd powers of
$\theta$, taking always into account the additional odd
contribution coming from the commutator of two operators.

In the commutative limit, since the terms

\beq [ e^a_\mu, \Lambda_0 ] \sim \{ f^a_\mu , \Lambda_5 \} \sim
\epsilon_{abcd} [ f^d_\mu, \Lambda^{bc} ] \sim O( \theta^2 ),
\label{45} \eeq

the usual transformation of the vierbein under local Lorentz
transformations is recovered

\beq \delta e^a_\mu = 2 \{ e^b_\mu, \Lambda^{ba} \} . \label{46}
\eeq

The vierbein transformations are diagonal at a level of the
combination $e^{\pm a}_\mu$, as already anticipated

\beq \delta e^{\pm a}_\mu = i [ e^{\pm a }_\mu, \Lambda_0 ] + 2 \{
e^{\pm b}_\mu, \Lambda^{ba} \} \pm i \{ e^{\pm a}_\mu, \Lambda_5
\} \pm i \epsilon_{abcd} [ e^{\pm d }_\mu, \Lambda^{bc} ] .
\label{47} \eeq

Analogously the spin connection transforms under $U(2) \otimes
U(2)$ according to the law

\beq X_\mu \rightarrow U^{\dagger} X_\mu U \label{48} \eeq

and at an infinitesimal level

\beq \delta X_\mu = [ X_\mu , \Lambda ] . \label{49} \eeq

In components this equation reads

\begin{eqnarray}
\delta \omega^1_\mu & = & i [ \hat{p}_\mu, \Lambda_0 ] + i [
\omega^1_\mu, \Lambda_0 ] + i [ \omega^5_\mu, \Lambda_5 ] + 2i [
\omega^{ab}_\mu, \Lambda^{ba} ] \nonumber \\
\delta \omega^5_\mu & = & i [ \hat{p}_\mu, \Lambda_5 ] + i [
\omega^1_\mu, \Lambda_5 ] + i [ \omega^5_\mu, \Lambda_0 ] + i
\epsilon_{abcd} [ \omega_\mu^{ab}, \Lambda^{cd} ] \nonumber \\
\delta \omega^{ab}_{\mu} & = & - i [ \hat{p}_\mu, \Lambda^{ab} ] -
i [ \omega^1_\mu, \Lambda^{ab} ] + \frac{i}{2} \epsilon_{abcd} [
\omega^5_\mu, \Lambda^{cd} ] + i  [ \omega^{ab}_{\mu}, \Lambda_0 ]
\nonumber \\
& - & \frac{i}{2} \epsilon_{abcd} [ \omega_\mu^{cd}, \Lambda_5 ] +
2 \{ \omega_\mu^{ac}, \Lambda^{cb} \} - 2 \{ \omega_\mu^{bc},
\Lambda^{ca} \} . \label{410}
\end{eqnarray}

The same considerations made for the vierbein apply here, i.e. the
spin connections $\omega^1_\mu$ and $\omega^5_\mu$,  restricted to
odd powers of $\theta$, and $\omega_\mu^{ab}$ restricted to even
powers of $\theta$, maintain this property if the gauge parameters
belong to $SO(4)_{*}$.

In the classical limit we notice that the only terms which survive
are all the expected ones:

\beq \delta \omega_\mu^{ab} = - i [ \hat{p}_\mu, \Lambda^{ab} ] +
2 \{ \omega_\mu^{ac}, \Lambda^{cb} \} - 2 \{ \omega_\mu^{bc},
\Lambda^{ca} \} . \label{411} \eeq

These properties can also be made more clear and transparent by
using the chiral decomposition for $X_\mu$:

\beq X_\mu = X_\mu^L \left( \frac{1+\gamma_5}{2}\right) + X_\mu^R
\left( \frac{1-\gamma_5}{2}\right) \label{412} \eeq

where

\beq X_\mu^L = \hat{p}_\mu + ( \omega^1_\mu + \omega^5_\mu ) + i
\omega^{ab}_\mu \gamma_{ab} \left( \frac{1+\gamma_5}{2}\right).
\label{413} \eeq

By using the property

\beq \gamma_{ab} \left( \frac{1+\gamma_5}{2} \right) = \half [
\gamma_{ab} - \half \epsilon_{abcd} \gamma_{cd} ] \label{414} \eeq

the $su(2)_{\rm \ left}$ part can be simplified to

\begin{eqnarray}
& \ & i \omega^{ab}_\mu \gamma_{ab} \left( \frac{1+\gamma^5}{2}
\right) = \frac{i}{2} ( \omega^{ab}_\mu - \half \omega^{cd}_\mu
\epsilon_{abcd} ) \gamma_{ab} \left( \frac{1+\gamma^5}{2} \right)
\nonumber \\
& \ & = i \tilde{\omega}^{L \ ab}_{\mu} \gamma_{ab} \left(
\frac{1+\gamma^5}{2} \right) . \label{415} \end{eqnarray}

It is not difficult to show that $\tilde{\omega}^{L \ ab}_\mu$ is
an anti-self-dual gauge connection:

\begin{eqnarray}
& \ & \tilde{\omega}^{L \ ab}_{\mu} = \half ( \omega^{ab}_\mu -
\half \omega^{cd}_\mu \epsilon_{abcd} ) \nonumber \\
& \ & \tilde{\omega}^{L \ ab}_{\mu} = - \half \epsilon^{abcd}
\tilde{\omega}^{L}_{\mu \ cd} . \label{416}
\end{eqnarray}

Analogously $X_\mu^R$ is made by a self-dual gauge connection:

\begin{eqnarray}
X_\mu^R & = & \hat{p}_\mu + ( \omega^1_\mu - \omega^5_\mu ) + i
\tilde{\omega}^{R \ ab }_{\mu} \gamma_{ab} \left( \frac{1-
\gamma_5}{2} \right) \nonumber \\
\tilde{\omega}^{R \ ab}_{\mu} & = & \half ( \omega^{ab}_\mu +
\half \omega^{cd}_\mu \epsilon_{abcd} ) \nonumber \\
\tilde{\omega}^{R \ ab}_{\mu} & = &  \half \epsilon^{abcd}
\tilde{\omega}^{R}_{\mu \ cd} . \label{417}
\end{eqnarray}

Reducing the gauge group to a pure left part by taking:

\beq U = U_L \left( \frac{1+\gamma_5}{2} \right) +  \left(
\frac{1-\gamma_5}{2} \right) \ \ \ \  \ \  U_R = 1 \label{418}
\eeq

then only the left part of the spin connection changes according
to

\begin{eqnarray}
X_\mu^L & \rightarrow & U^{\dagger}_L X_\mu^L U_L \nonumber \\
X_\mu^R & \rightarrow & X_\mu^R . \label{419}
\end{eqnarray}

The implications of this observation are interesting, since
applying a pure left unitary transformation to the background
connection $ X_\mu = \hat{p}_\mu$ one constructs a ( pure gauge )
anti-self-dual spin connection, while $X_\mu^R$ remains pure
background.

Decomposing the gauge parameter $\Lambda$ into chiral components
leads to:

\beq \Lambda = \Lambda^L \left( \frac{1+\gamma_5}{2} \right) +
\Lambda^R \left( \frac{1-\gamma_5}{2} \right) \label{420} \eeq

where

\begin{eqnarray}
\Lambda^L & = & i \Lambda^0_L + \tilde{\Lambda}^{ab}_L \gamma_{ab}
\left( \frac{1+\gamma_5}{2} \right) \ \ \  \ \ \ \
\tilde{\Lambda}^{ab}_L = \half ( \Lambda^{ab} - \half
\epsilon^{abcd} \Lambda_{cd} ) \nonumber \\
\Lambda^R & = & i \Lambda^0_R + \tilde{\Lambda}^{ab}_R \gamma_{ab}
\left( \frac{1-\gamma_5}{2} \right) \ \ \  \ \ \ \
\tilde{\Lambda}^{ab}_R = \half ( \Lambda^{ab} + \half
\epsilon^{abcd} \Lambda_{cd} ) \nonumber \\
\Lambda^0_L & = & \Lambda_0 + \Lambda_5 \ \ \ \ \ \ \ \Lambda^0_R
= \Lambda_0 - \Lambda_5 . \label{421}
\end{eqnarray}

Defining $ \omega^L_\mu = \omega^1_\mu + \omega^5_\mu$,
$\omega^R_\mu = \omega^1_\mu - \omega^5_\mu$, $X_\mu^L$ can be
rewritten as:

\beq X^L_\mu = \hat{p}_\mu + \omega_\mu^L + i
\tilde{\omega}_{\mu}^{ab} \gamma_{ab} \left( \frac{1+\gamma_5}{2}
\right) \label{422} \eeq

and the gauge property of $X_\mu^L$ now reads:

\begin{eqnarray}
\delta X_\mu^L & = & [ X_\mu^L, \Lambda^L ] \nonumber \\
\delta \omega^L_\mu & = & i [ \hat{p}_\mu, \Lambda^0_L ] + i [
\omega_\mu^L, \Lambda^0_L ] + 4i [ \tilde{\omega}^{L \ ab }_\mu,
\tilde{\Lambda}^{ba}_L ] \nonumber \\
\delta \tilde{\omega}^{L \ ab}_\mu & = & - i [ \hat{p}_\mu,
\tilde{\Lambda}^{L \ ab } ] - i [ \omega^L_\mu, \tilde{\Lambda}^{L
 \ ab } ] + i [ \tilde{\omega}^{L \ ab}_\mu, \Lambda^0_L ]
 \nonumber \\
 & + & 2 \{ \tilde{\omega}^{L \ ac}_\mu, \tilde{\Lambda}^{cb}_L \}
- 2 \{ \tilde{\omega}^{L \ bc}_\mu, \tilde{\Lambda}^{ca}_L \} .
\label{423}
\end{eqnarray}

In the classical limit the anti-self-dual spin connection
$\tilde{\omega}^{L \ ab}_\mu$ transforms under the anti-self-dual
gauge parameter $\tilde{\Lambda}^{ab}_L$.

\section{Definition of the metric}

It has been already pointed out in the literature that the metric
is given by the star product of two vierbeins and it is no more
symmetric, but these observations are not conclusive in my
opinion. There is one more difficulty to set up a consistent
second order formalism for noncommutative gravity, i.e. that the
metric tensor is not even invariant under the gauge group $U(2)
\otimes U(2)$, on which the model is defined.

The only way out to this further obstacle is to allow for a more
general definition of metric, as a bilinear combination of the
vierbein which is at least covariant under $U(2) \otimes U(2)$,
i.e. a representation of the basic gauge group of the theory.

We have in fact at disposition the bilinear form

\beq Y_\mu Y_\nu = G_{\mu\nu} + i B_{\mu\nu} \ \ \ \  G_{\mu\nu} =
\half \{ Y_\mu, Y_\nu \} \ \ \ \ B_{\mu\nu} = - \frac{i}{2} [
Y_\mu, Y_\nu ] \label{51} \eeq

which transforms in a covariant way under $U(2) \otimes U(2)$,
having the same transformation properties of the spin connection
$X_\mu$ ( apart from the presence of two world indices instead of
one ).

Decomposing (\ref{51}) into chiral parts one finds:

\beq Y_\mu Y_\nu = Y^{+}_\mu Y^{-}_\nu \left( \frac{1-\gamma_5}{2}
\right) + Y^{-}_\mu Y^{+}_\nu \left( \frac{1+\gamma_5}{2} \right).
\label{52} \eeq

Therefore we conclude that it is not possible to define a
covariant metric tensor with only one vierbein $e^{+ a}_\mu$, but
it is necessary the presence of both vierbeins.

Expanding the bilinear form (\ref{51}) into components one finds:

\beq Y_\mu Y_\nu = Y_{\mu\nu}^0 + Y_{\mu\nu}^5 \gamma_5  +
Y_{\mu\nu}^{ab} \gamma_{ab} \label{53} \eeq

with

\begin{eqnarray}
Y^0_{\mu\nu} & = & \eta_{ab} ( e^a_\mu e^b_\nu - f^a_\mu f^b_\nu )
\nonumber \\
Y^5_{\mu\nu} & = & i \eta_{ab} ( e^a_\mu f^b_\nu - f^a_\mu e^b_\nu
) \nonumber \\
Y^{ab}_{\mu\nu} & =  & \half ( e^{[ \ a}_\mu e^{b \ ]}_\nu - f^{[
\ a}_\mu f^{b \ ]}_\nu ) - \frac{i}{2} \epsilon_{abcd} ( e^c_\mu
f^d_\nu - f^c_\mu e^d_\nu ) \label{54}
\end{eqnarray}

where the symbols between parenthesis [ \ ] mean that we must take
the antisymmetric combination of the indices.

The tensor $G_{\mu\nu}$ is  symmetric and hermitian and its
components are given by

\begin{eqnarray}
G_{\mu\nu} & = & g_{\mu\nu}^0 + g_{\mu\nu}^5 \gamma_5 +
g_{\mu\nu}^{ab} \gamma_{ab} = \nonumber \\
& = & \half [ ( \{ e^a_\mu, e^b_\nu \} + \{ f^a_\mu, f^b_\nu \}
\eta_{ab} + i ( [ e^a_\mu, f^b_\nu ] + [ e^a_\nu, f^b_\nu ] )
\eta_{ab} \gamma_5 \nonumber \\
& + & ( [ e^a_\mu, e^b_\nu ] + [ f^a_\mu, f^b_\nu ] - \frac{i}{2}
\epsilon^{abcd} ( \{ e^c_\mu, f^d_\nu \} + \{ e^c_\nu  f^d_\mu \}
) ) \gamma_{ab} ] . \label{55}
\end{eqnarray}

The $g_{\mu\nu}^0$ and $g_{\mu\nu}^5$ parts can be restricted to
contain only even powers of $\theta$, while $g_{\mu\nu}^{ab}$
contains only odd powers of $\theta$, and since obviously

\beq g_{\mu\nu}^5 \sim O(\theta^2) \label{56} \eeq

the only part which survives the commutative limit is

\beq g^0_{\mu\nu} = \half \{ e^a_\mu, e^b_\nu \} \eta_{ab} .
\label{57} \eeq

Under a gauge transformation, $G_{\mu\nu}$ transforms as

\beq G_{\mu\nu} \rightarrow U^{\dagger} G_{\mu\nu} U \label{58}
\eeq

or at an infinitesimal level

\beq \delta G_{\mu\nu} = [ G_{\mu\nu}, \Lambda ] . \label{59} \eeq

In components one finds

\begin{eqnarray}
\delta g_{\mu\nu}^0 & = & i [ g_{\mu\nu}^0, \Lambda_0 ] + i [
g_{\mu\nu}^5, \Lambda^5 ] + 2i [ g^{ab}_{\mu\nu}, \Lambda^{ba} ]
\nonumber  \\
\delta g_{\mu\nu}^5 & = & i [ g_{\mu\nu}^0, \Lambda^5 ] + i [
g^5_{\mu\nu}, \Lambda^0 ] + i \epsilon_{abcd} [ g^{ab}_{\mu\nu},
\Lambda^{cd} ] \nonumber \\
\delta g_{\mu\nu}^{ab} & = & -i [ g^0_{\mu\nu}, \Lambda^{ab} ] +
\frac{i}{2} \epsilon_{abcd} [ g^5_{\mu\nu}, \Lambda^{cd} ] + i [
g^{ab}_{\mu\nu}, \Lambda_0 ] \nonumber \\
& - & \frac{i}{2} \epsilon_{abcd} [ g^{cd}_{\mu\nu}, \Lambda_5 ] +
2 \{ g^{ac}_{\mu\nu}, \Lambda^{cb} \} - 2 \{ g_{\mu\nu}^{bc},
\Lambda^{ca} \} . \label{510}
\end{eqnarray}

It is clear that $\delta g^0_{\mu\nu} \sim O(\theta^2)$, therefore
one reobtains that $g^0_{\mu\nu}$ is gauge invariant in the
classical limit.

For the antisymmetric and hermitian part $B_{\mu\nu}$ one finds
analogous gauge transformations properties, and in this case the
components $B_{\mu\nu}^0$, $B_{\mu\nu}^5$ can be restricted to
contain only odd powers of $\theta$, while $B^{ab}_{\mu\nu}$ only
even powers of $\theta$. In the classical limit there is one term
which survives

\beq B_{\mu\nu}^{ab} = - \frac{i}{4} ( \{ e^a_\mu, e^b_\nu \} - (
a \leftrightarrow b ) ) \label{511} \eeq

and it transforms in a covariant way as we can see from the
formula (\ref{510}). This antisymmetric part however decouples
from the Einstein equations, and it can be neglected.

In conclusion, in order to setup a consistent second order
formalism we believe that it is necessary to include all these
components into the game. Some results contained in \cite{12}
confirm indirectly this picture.

\section{Gravitational instantons}

Given a solution of the equations of motion (\ref{32}) and
(\ref{35}), it is possible to generate another one which is not
trivially connected to the first one by introducing a
quasi-unitary operator, which in the case of $U(2) \otimes U(2)$
is of the type

\begin{eqnarray}
& \ & U U^{\dagger} = 1   \ \ \ \ \ \ \ \ U^{\dagger} U = 1 - P_0
\nonumber \\
& \ & X_\mu \rightarrow U^{\dagger} X_{\mu} U \ \ \ \ \ \ \ \ \
Y_{\mu} \rightarrow U^{\dagger} Y_{\mu} U \label{61}
\end{eqnarray}

where $P_0$ is a projector with values in $U(2) \otimes U(2)$.

In particular we can start from the vacuum, which is defined by
the background of the matrix model:

\begin{eqnarray}
X_\mu & = & \hat{p}_\mu \nonumber \\
Y_\mu & = & \delta^a_\mu \gamma_a \label{62}
\end{eqnarray}

and compute the following transformations

\begin{eqnarray}
X_\mu & = & U^{\dagger} \hat{p}_\mu U \nonumber \\
Y_\mu & = & U^{\dagger} \delta^a_\mu \gamma_a U . \label{63}
\end{eqnarray}

These are automatically solutions to the equations of motion of
the matrix model (\ref{32}) and (\ref{35}), due to the property $U
U^{\dagger} = 1$, and give a finite contribution to the matrix
model action since then:

\beq [ X_\mu, X_\nu ] - i \theta^{-1}_{\mu\nu} = i
\theta^{-1}_{\mu\nu} P_0 \label{64} \eeq

this commutator is a projector, and the trace defining the action
is projected on a finite number of states. We call this generic
solution a noncommutative gravitational instanton.

Obviously one can introduce more structure into the game, by
requiring that the noncommutative solutions have a smooth
$\theta$-limit and coincide in the commutative limit with some
known and classified solution \cite{13}-\cite{17}. It is not the
purpose of the present paper, however we believe that our
definition can be adjusted to achieve all these goals. For the
moment we limit ourself to indicate some general property of our
finite action solutions of the equations of motion.

Firstly the projector $P_0$ can be decomposed into chiral parts

\beq P_0 = P_0^L \left( \frac{1+\gamma_5}{2} \right) + P_0^R
\left( \frac{1-\gamma_5}{2} \right) \label{65} \eeq

where $P_0^L$ and $P_0^R$ are two independent $U(2)$-valued
projectors. Let us note that in the particular case in which one
of these two projectors is null, the corresponding quasi-unitary
operator is of the form:

\beq U = U_L \left( \frac{1+\gamma_5}{2} \right) +  \left(
\frac{1-\gamma_5}{2} \right) \ \ \ \ {\rm or } \ \ \ U =  \left(
\frac{1+\gamma_5}{2} \right) + U_R \left( \frac{1-\gamma_5}{2}
\right) \label{66} \eeq

and it produces a nontrivial spin connection only for the left or
right sector:

\beq X_\mu = U^{\dagger}_L \hat{p}_\mu U_L \left(
\frac{1+\gamma_5}{2} \right) + \hat{p}_\mu  \left(
\frac{1-\gamma_5}{2} \right) \label{67} \eeq

or

\beq X_\mu =  \hat{p}_\mu  \left( \frac{1+\gamma_5}{2} \right) +
U^{\dagger}_R \hat{p}_\mu  U_R \left( \frac{1-\gamma_5}{2} \right)
. \label{68} \eeq

The corresponding spin connection is automatically anti-self-dual
or self-dual, according to the choice $P_0^R = 0 $ or $P_0^L = 0
$.

Therefore solutions with self-dual or anti-self-dual spin
connections are naturally implemented, although the class of
solution defined by (\ref{65}) is more general.

What is the generic form of the projector $P_0^L$ with values in
$U(2)_{\rm \ left}$ ? We don't have the general proof but we
believe that the more general solution is of the form

\beq P_0^L = P_0^{+} \left( \frac{1+\sigma_3}{2} \right) + P_0^{-}
\left( \frac{1-\sigma_3}{2} \right) \label{69} \eeq

where $P_0^{\pm}$ are two independent projectors of $U(1)$, apart
from eventual isomorphisms of the Hilbert space, which are
implemented by unitary transformations of the type

\beq P_0 \rightarrow U^{\dagger} P_0 U \ \ \ \ \  \ U^{\dagger} U
= U U^{\dagger } = 1 . \label{610} \eeq

We can therefore reduce the general problem of projectors with
values in $U(2) \otimes U(2)$ to the $U(1)$ case.

This is also clear from the duality between $U(2)$ and $U(1)$
gauge groups on the noncommutative plane which can be described as
follows.

Consider the Hilbert space created by the commutation rules of the
coordinates and label it with a quantum number $n$. The
correspondence between $U(2)$ and $U(1)$ is obtained by, firstly
enlarging the one-oscillator basis to the $U(2)$ basis

\beq |n; a> \ \forall n \in N \ \ a = 0,1 \label{611} \eeq

and noting the isomorphism between Hilbert spaces:

\beq {\cal H } \rightarrow {\cal H} \ \ \ \ \  |n: a > \rightarrow
| 2n+a > \ \ \ \ \ a = 0,1  .\label{612} \eeq

Therefore we can relabel the tensorial product of the Hilbert
space of one oscillator and the gauge group $U(2)$ with a new
quantum number $n' = 2n + a$ that describes only oscillator states
with $U(1)$ gauge group \cite{18}.

Once that the general $U(2) \otimes U(2)$ projector is reduced to
a $U(1)$ projector, we need to build quasi-unitary operators with
gauge group $U(1)$ in four dimensions.

In four dimensions the Hilbert space on which the quasi-unitary
operator acts is ( see Ref. \cite{18} ) generally the tensorial
product of two Hilbert spaces of one oscillator

\beq {\cal H} \times {\cal H} \ \ \ \ |n_1,n_2> \ \ \forall \ n_1,
n_2 \in N . \label{613} \eeq

We can therefore introduce a duality between four dimensions and
two dimensions, by observing that a couple of numbers can be made
isomorphic to a number, for example

\beq (n_1, n_2) \rightarrow \frac{(n_1+n_2)(n_1+n_2+1)}{2} + n_2 \
\ \ \ \ \ \ {\cal H} \times  {\cal H} \rightarrow {\cal H}
\label{614} \eeq

and therefore we can relabel the Hilbert space with only one
quantum number.

Explicitly the construction of a general class of quasi-unitary
operators with values in $U(1)$ in four dimensions follows these
steps. Let us define two new quantum numbers

\beq n = n_1 + n_2 \ \ \ \ \ k= n_2 \label{615} \eeq

and let us introduce a short notation for the state:

\beq | n_1, n_2 > \equiv | \frac{n(n+1)}{2} + k > = | n; k > .
\label{616} \eeq

A basis of the two-dimensional Hilbert space is determined by the
states

\beq |n;k> \ \ \ \ 0 \leq k \leq n \ \ \ \ \forall n \in N .
\label{617} \eeq

We must allow the continuation of the notation (\ref{617}) to
states with $k \geq n$ keeping in mind the following equivalence
relation:

\beq |n;k > = | n+1; k-n-1 > . \label{618} \eeq

In the two-dimensional basis, the generic finite projector
operator $P_0$ can be represented in the following form, apart
from an isomorphism of the Hilbert space,

\beq P_0 = \sum_{i=0}^{m-1} |i ><i| \label{619} \eeq

that represents a configuration with instanton number $m$.

In the two-dimensional basis it is easy to derive the
quasi-unitary operator $U$ that produces the projector operator
$P_0$:

\begin{eqnarray}
U U^{\dagger} & = & 1 \ \ \ \ \  U^{\dagger} U = 1 - P_0 \nonumber \\
U & = & \sum^{\infty}_{n=0} \sum^{n}_{k=0} |n;k ><n; k+m |
\nonumber \\
U^{\dagger} & = & \sum^{\infty}_{n=0} \sum^{n}_{k=0} |n;k+m ><n; k
| . \label{620}
\end{eqnarray}

To derive the quasi-unitary operator in the equivalent basis
$|n_1,n_2 >$ we must pullback the duality from the
four-dimensional plane and the two-dimensional one. The problem is
complicated in general, and it is simple to perform it only in the
simplest case, with a configuration with instanton number $m$.
Then the quasi-unitary operator can be reexpressed as:

\begin{eqnarray}
U & = & \sum^{\infty}_{n_1=0} \sum^{\infty}_{n_2=0} | n_1 +1, n_2
>< n_1 , n_2 + 1 | \nonumber \\
& + & \sum_{n_1=n_2=0}^{\infty} |0, n_2 >< n_1+1, 0 | \nonumber \\
U^{\dagger} & = & \sum^{\infty}_{n_1=0} \sum^{\infty}_{n_2=0} |
n_1 , n_2 + 1
>< n_1 + 1  , n_2 | \nonumber \\
& + & \sum_{n_1=n_2=0}^{\infty} |n_1+1 , 0 >< 0, n_2 | .
\label{621}
\end{eqnarray}

In summary, the construction of gravitational instantons can be
derived as in the Yang-Mills case by introducing quasi-unitary
operators with values in $U(2) \otimes U(2)$. Self-dual or
anti-self-dual solutions can be achieved restricting the gauge
group to its left part or right part only, being the general
solution defined by eqs. (\ref{65}) without any particular
symmetry. The question of the smoothness of the classical limit
remains to be investigated, as well as the ( noncommutative )
characterization of these solutions in terms of topological
invariants.

\section{Conclusions}

In this paper we have attempted to define noncommutative gravity
theory with a Matrix model approach. The noncommutative plane is
taken as a background solution of the Matrix model, and the
fluctuations are the vierbein and spin connections.

Two types of vierbein are needed to make the formalism consistent
at the noncommutative level. It is not possible to make them
proportional, but it is possible to restrict the equations of
motion in such a way that one type of vierbein has only even
powers of $\theta$, and the other one only odd powers of $\theta$,
recovering in the classical limit the usual gravity theory with
only one vierbein.

The spin connection has other two $U(1)$ parts, which however can
be restricted to contain only odd powers of $\theta$ and therefore
are negligible in the $\theta \rightarrow 0$ limit.

These properties can be respected by gauge transformations, if the
gauge parameters are restricted to the $SO(4)_{*}$, the smallest
subgroup of $U(2) \otimes U(2)$ , consistent with the star
product. The distinction between odd and even powers of $\theta$
is fruitful also in the discussion of the $\theta \rightarrow 0$
limit of the equations of motion, in which we recover the Einstein
equations.

We have then attacked the problem of defining a consistent second
order formalism. The metric tensor which is bilinear in the
vierbein cannot be defined to be invariant under $U(2) \otimes
U(2)$ or even $SO(4)_{*}$ in the noncommutative case, but only
covariant. The need of a multiplet of metric tensors, each one
having no direct physical meaning, is confirmed indirectly by the
computations of ref. \cite{12}.

Finally we have attempted to give a definition of noncommutative
gravitational instantons. We have introduced $U(2) \otimes U(2)$
valued quasi-unitary operators which generate, once that are
applied to the background, nontrivial solutions of the equations
of motion.

In the case of a pure left or right quasi-unitary operators, the
corresponding solution has a self-dual or anti-self-dual spin
connection, a property which defines the commutative gravitational
instantons. Our class of finite action solutions is more general.
We have then constructed explicit examples of quasi-unitary
operators based on the concept of duality of Hilbert spaces, which
is typical of the noncommutative plane.

Finally let us briefly mention the problems left; firstly the
construction of a consistent second-order formalism, with the use
of the multiplet of metric tensors outlined in this paper,
secondly the careful analysis of the physical degrees of freedom
of the metric tensor, and the cancellation of the unphysical ones
and thirdly a more careful analysis of the noncommutative
gravitational instantons and of their link with the commutative
case. The last project will require to take control of the $\theta
\rightarrow 0$ limit of the nonperturbative solutions constructed
in this article, and the introduction of topological invariants
for the noncommutative case. In any case we believe that the
language of matrices, introduced in this paper, which is more
familiar to a physicist than the star product formalism, will help
in making progress in this research field.

\appendix
\section{Appendix}

The gamma matrices are known to generate the $U(4)$ and $U(2,2)$
algebra. In this appendix we recall some basic properties, like
the basic representations and commutation properties which are
used during this paper.

Firstly we repeat the $U(4)$ ( Euclidean ) case. We must solve the
anticommutation relations

\beq \{ \gamma_a, \gamma_b \} = 2 \delta_{ab} \label{A1} \eeq

with the constraints $\gamma^{\dagger}_a = \gamma_a$. The solution
 to this requirements is (  the so-called chiral representation )

 \beq \gamma_0 = \left( \begin{array}{cc} 0 & 1 \\ 1 & 0
 \end{array} \right) \ \ \ \ \ \ \gamma_i = \left( \begin{array}{cc} 0 & i \sigma_i
 \\ - i \sigma_i  & 0 \end{array} \right) \label{A2}
 \eeq

 where $\sigma_i$ are the Pauli matrices. The properties

 \beq \gamma^2_5 = 1 \ \ \ \ \gamma^{\dagger}_5 = \gamma_5 \label{A3} \eeq

 identify $\gamma_5$ as the combination

 \beq \gamma_5 = \gamma_0 \gamma_1 \gamma_2 \gamma_3 \ \ \ \ \ \
 \gamma_5 = \left( \begin{array}{cc} 1 & 0 \\ 0 & -1 \end{array}
 \right) . \label{A4} \eeq

 In the case of $U(2,2)$, we have to solve the anticommutation
 relations:

 \beq \{ \gamma_a, \gamma_b \} = 2 \eta_{ab} \ \ \ \ \ \ ( + + - -
 ) \label{A5} \eeq
 with the constraints

 \beq \gamma^{\dagger}_0 = \gamma_0 \ \ \ \gamma^{\dagger}_1 = \gamma_1 \ \ \
 \gamma^{\dagger}_2 = - \gamma_2 \ \ \gamma^{\dagger}_3 = -
 \gamma_3.
 \label{A6} \eeq

 A possible choice is , in the chiral representation

 \beq \gamma_0 = \left( \begin{array}{cc} 0 & 1 \\ 1 & 0
 \end{array} \right) \ \ \
\gamma_1 = \left( \begin{array}{cc} 0 & -i \sigma_1 \\ i \sigma_1
& 0 \end{array} \right) \ \ \
\gamma_2 = \left( \begin{array}{cc} 0 & - \sigma_2 \\
\sigma_2 & 0 \end{array} \right) \ \ \ \gamma_3 =
\left(\begin{array}{cc} 0 & -\sigma_3 \\ \sigma_3 & 0
 \end{array} \right) . \label{A7} \eeq

 Again $\gamma_5$ identified with the properties (\ref{A3}) can be chosen
 as in the Euclidean case (\ref{A4}).

 The hermitian conjugation property can be encoded in the
 following property

 \beq \gamma_a^{\dagger} = \Gamma_0 \gamma_a \Gamma_0 \label{A8} \eeq

 where $\Gamma_0 = \gamma_0 \gamma_1 $.

 The composition properties of gamma matrices can be summarized as
 follows:

 \begin{eqnarray}
 & \ & [ \gamma_a , \gamma_{bc} ] = 2 \eta_{ab} \gamma_c - 2
 \eta_{ac} \gamma_b \nonumber \\
 & \ & \{ \gamma_a , \gamma_{bc} \} = 2 \epsilon_{abcd} \gamma_5
 \gamma^d \nonumber \\
 & \ & [ \gamma_{ab}, \gamma_{cd} ] = 2 ( \eta_{ad} \gamma_{bc} +
 \eta_{bc} \gamma_{ad} - \eta_{ac} \gamma_{bd} - \eta_{bd}
 \gamma_{ac} ) \nonumber \\
 & \ & \{ \gamma_{ab}, \gamma_{cd} \} = 2 ( \eta_{ad} \eta_{bc} -
 \eta_{ac} \eta_{bd} ) + 2 \epsilon_{abcd} \gamma_5 \nonumber \\
 & \ & \gamma_5 \gamma_{ab} = - \half \epsilon_{abcd} \gamma^{cd}
 \label{A9}
 \end{eqnarray}

 where we raise the indices with the tensor $\eta^{ab}$. Of course
 in the Euclidean case the distinction between upper and lower
 indices is superfluous.

\end{document}